\documentclass[useAMS,usenatbib]{mn2e}
\usepackage{graphicx}
\usepackage{epsfig}

\title[{\it Suzaku} Observations of Ejecta-Dominated G346.6-0.2]{{\it Suzaku} Observations of Ejecta-Dominated
Galactic Supernova Remnant G346.6-0.2}
\author[A.~Sezer, F.~G\"{o}k, M.~Hudaverdi, M.~Kimura and E.N.~Ercan]{A.~Sezer,$^{1,2}$\thanks{E-mail: aytap.sezer@uzay.tubitak.gov.tr
(AS); gok@akdeniz.edu.tr (FG); murat.hudaverdi@uzay.tubitak.gov.tr
(MH); mkimura@ess.sci.osaka-u.ac.jp (MK); ercan@boun.edu.tr
(ENE).} F.~G\"{o}k,$^{3}$ M.~Hudaverdi,$^{1}$ M.~Kimura$^{4}$ and
E.N.~Ercan$^{2}$\footnotemark[1]\thanks{This file has been amended
to highlight the proper use of \LaTeXe\ code with the class file.
These changes are for illustrative purposes and do not reflect the
original paper by A. Sezer.}\\
$^{1}$T\"UB\.ITAK Space Technologies Research Institute, ODTU
Campus, Ankara, 06531, Turkey\\
$^{2}$Bo\~gazi\d{c}i University, Faculty of Art and Sciences,
Department of Physics, \.Istanbul, 34342, Turkey\\
$^{3}$Akdeniz University, Faculty of Sciences, Department
of Physics, Antalya, 07058, Turkey\\
 $^{4}$Osaka University, Department of Earth and Space Science,
Graduate School of Science,\\
Machikaneyama 1-1, Toyonaka, Osaka,
560-0043, Japan\\
}
\begin{document}

\date{}

\pagerange{\pageref{firstpage}--\pageref{lastpage}} \pubyear{2002}

\maketitle

\label{firstpage}

\begin{abstract}
We present here the results of the X-ray analysis of Galactic
supernova remnant G346.6-0.2 observed with {\it Suzaku}. K-shell
emission lines of Mg, Si, S, Ca and Fe are detected clearly for
the first time. Strong emission lines of Si and S imply that X-ray
emission nature of G346.6-0.2 is ejecta-dominated. The
ejecta-dominated emission is well fitted with a combined model
consisting of thermal plasma in non-equilibrium ionization and a
non-thermal component, which can be regarded as synchrotron
emission with a photon index of $\Gamma$ $\sim 0.6$. Absorbing
column density of $N_{\rm H}\sim2.1\times10^{22}$ ${\rm cm^{-2}}$
is obtained from the best-fitting implying a high-density medium,
high electron temperature of $kT_{\rm e}\sim1.2$ keV, and
ionization timescale of $n_{\rm e}t\sim2.9\times10^{11}$ ${\rm
cm^{-3}s}$ indicating that this remnant may be far from full
ionization equilibrium. The relative abundances from the ejecta
show that the remnant originates from a Type Ia supernova
explosion.
\end{abstract}


\section{Introduction}
A supernova remnant (SNR) consists of expelled material called
``ejecta'' from the explosion and a swept-up interstellar matter.
The X-ray emission results from the interactions of ejecta and
swept-up matter. From the X-ray emission of SNRs we may obtain
valuable information about the physical properties of the ejecta,
swept-up plasma, elemental abundances and the history of the
explosion. Shock wave may be the blast wave associated with the
stellar explosion and/or the reverse shock wave, which propagates
inwards from the decelerated blast wave and raises the temperature
of the stellar ejecta. The young SNRs are bright in X-rays and
dominated by the emission from ejecta. Thus they provide fruitful
information about the elements synthesized by the supernova (SN)
explosions. Therefore, the observation of the young SNRs is the
best method to investigate the abundances of the elements
synthesized by the SNe (see e.g. SN1006 \citep {b11}, RCW86 \citep
{b3}, Tycho \citep {b24}  for the best studied ejecta-dominated
SNRs in the X-ray band).

G346.6-0.2 ($\rmn{RA}(2000)=17^{\rmn{h}} 10^{\rmn{m}}
19^{\rmn{s}}$, $\rmn{Dec.}~(2000)=-40\degr 11\arcmin$) is a
shell-type SNR located in the Galactic plane. It was discovered by
\citet {b5} in radio band having an angular size of 8 arcmin
\citep {b25}. In X-ray band, on the other hand, G346.6-0.2 was
first observed by {\it ASCA} during its Galactic plane survey
\citep {b28}. It is shown that the size of the X-ray emission from
G346.6-0.2 is less extended than its reported radio structure.
Five OH(1720 MHz) masers were detected toward this SNR and they
are all located along the southern edge of the remnant \citep
{b10}.

We are studying the X-ray emissions from the ejecta-dominated
SNRs. For this purpose we have chosen several small size Galactic
SNRs, one of which is G346.6-0.2. Detailed properties of
G346.6-0.2 remained unknown so far because of very limited photon
statistics in the {\it ASCA} ({\it AGPS}) data. {\it Suzaku} is
the most recent X-ray astronomical satellite (see \citet {b17})
having a large collecting area and low background. Therefore, it
is the best instrument for observing dim and diffuse sources. We
propose here to study G346.6-0.2 to understand the origin of its
thermal and non-thermal X-ray emission which will help us to
distinguish its SN explosion type. By using the archival data of
{\it Suzaku}, we were able to produce higher quality image and the
spectra of the remnant, which lead to the results in this study.

In Section 2, we describe the observation log and the data
reduction methods. We present the image analysis in Section 2.1
and the spectral analysis in Section 2.2. We discuss our results
and the origin of the thermal and non-thermal emission in Section
3.

\section[]{Observation and Data Reduction}

{\it Suzaku} satellite has two sets of instruments, one being the
four X-ray imaging spectrometers (XISs, see \citet {b12}); each of
four are  at the focus of an X-ray telescope (XRTs, see \citet
{b21}) and also a separate hard X-ray detector (HXD, see \citet
{b22}; \citet {b9}). XIS's have two different type of CCDs: one
being three front illuminated CCDs (namely, FI, XIS0, XIS2 and
XIS3) and other is a back illuminated CCD (namely, BI, XIS1). Each
CCD covers an area of 17.8$\times$17.8 arcmin$^{2}$.

The data used in this analysis is taken by the XIS onboard {\it
Suzaku}. G346.6-0.2 was observed on 2009 October 07 for 56.7 ks
(Obs ID:504096010). During the observation, the XIS were operated
in the normal full-frame clocking mode with the editing mode of
$3\times3$ and $5\times5$, which are low and medium data rates,
high and super-high data rates, respectively.

For the data reduction and spectral analysis, we used the {\sc
headas} software package of version 6.5 and {\sc xspec} version
11.3.2 (see \citet {b2}). The response matrix files (RMF) and
ancillary response files (ARF) were made using {\sc xisrmfgen} and
{\sc xissimarfgen} version 2006-10-17 (see \citet {b7}).

\subsection{Image Analysis}
Figure 1 shows the XIS0 image of G346.6-0.2 in the $0.3-10$ keV
energy band. The solid circle (radius $\sim$5 arcmin) from the
centre) shows the region where the spectra is extracted. The
background region is shown by dotted circle
($\rmn{RA}(2000)=17^{\rmn{h}} 10^{\rmn{m}} 54^{\rmn{s}}$,
$\rmn{Dec.}~(2000)=-40\degr 06\arcmin 24\arcsec$, radius $\sim$1.8
arcmin). We excluded the upper right corner of the detectors
containing the onboard masked out $^{55}$Fe calibration sources.
For comparison radio continuum image of G346.6-0.2 at 843 MHz by
\citet{b25} is overlaid on the figure, where four maser sources
are also pointed out as has been reported by \citet {b10}.

The bright image in XIS0 is clearly seen and we have obtained the
Si-S map in $1.5-2.5$ keV energy band (Figure 2) to see if this
emission is from interstellar medium (ISM) or from the ejecta. In
Fig. 2, Si-S emission is distributed throughout the remnant and it
is bright, which may well indicate that the X-ray emission coming
from the remnant is ejecta dominated rather than originating from
the ISM.

\subsection{Spectral Analysis}

The extracted XIS spectra from a circular region of G346.6-0.2
(see Fig. 1) are analyzed in the full energy band of $0.3-10$ keV.
Figure 3 shows representative spectra of XIS0, XIS1 and XIS3
simultaneously, extracted from the region (shown in Fig. 1 by the
solid circle) with its corresponding best-fitting model and
residuals.

The strong Si and S lines from the spectra and Si-S image (see
Fig. 2) in $1.5-2.5$ keV energy band suggest that the emission is
ejecta dominated. Therefore, we have applied one-component models
in our spectral fits. We first fitted the data with an absorbed
(wabs; \citet {b18}) VNEI, which is a model for a non-equilibrium
ionization (NEI) collisional plasma with variable abundances
\citep {b4}. The model did not fit well (reduced $\chi^{2}$ of
1.12 (637/567 d.o.f.)). So, we added a power-law model
(non-thermal synchrotron emission from relativistic electrons).
The parameters of the absorbing column density ($N_{\rm H}$);
electron temperature ($kT_{\rm e}$); the ionization parameter
($n_{\rm e}t$) and the metal abundances C, N, O, Ne, Mg, Si, S,
Ar, Ca, and Fe are set free in this fit resulting in an
unacceptable error value. Then, we set Mg, S, Si, Ca, and Fe free,
since their emission lines are clearly seen in the spectra, while
the rest are set to solar values \citep{b1} resulting in a
reasonably good reduced $\chi^{2}$ value of 1.07 (606/565 d.o.f.)
with reasonable error limits. Similar steps were also applied to
VMEKAL model, which describes an emission spectrum from hot
ionized gas in collisional ionization equilibrium with variable
abundances \citep {b15, b16, b13}, and VPSHOCK model, which is
suitable for modelling plane-parallel shocks in young SNRs, where
plasma has not reached the ionization equilibrium \citep {b4}.
These two models gave us reduced $\chi^{2}$ value of 1.22 (691/568
d.o.f.) and 1.11 (629/567 d.o.f.), respectively. Then we added a
power-law component to both models, which yielded almost similar
results (of reduced $\chi^{2}/$d.o.f. value of 607/566 and
605/565, respectively, of the best-fitting parameters and errors
values) as VNEI model as given in Table 1.

These three models basically represent the emission from optically
thin thermal plasma with slight differences. For example, VNEI is
a NEI model similar to VPSHOCK, except with a single ionization
timescale.

From an absorbed VNEI and power-law model, the absorbing column
density, the electron temperature, the ionization parameter, the
metal abundances of Mg, Si, S, Ca, Fe, volume emission measure
(VEM) and flux ($F_{\rm x}$) of the VNEI component and the photon
index ($\Gamma$) and norm of power-law component are obtained and
given in Table 1, with the corresponding error values with 90 per
cent confidence levels. Total flux and reduced $\chi^{2}/$d.o.f.
value of VNEI and power-law components are also given in this
table. The results of VMEKAL and power-law, VPSHOCK and power-law
models are also presented in Table 1 for comparison.

To determined the line centre energy of the K-shell ($K\alpha$)
lines, we fitted the spectra using a simple bremsstrahlung
continuum plus five Gaussian lines with an absorbing column. The
obtained best-fitting central energies of $K\alpha$ emission lines
are He-like Mg (1.38$\pm0.06$ keV), Si (1.86$\pm0.02$ keV), S
(2.52$\pm0.06$ keV), Ca (3.25$\pm0.55$ keV) and Fe (6.77$\pm0.32$
keV).

Figure 4 shows the best-fitting metal abundances (normalized to
Si) relative to solar values \citep{b1} with the predicted
nucleosynthesis W7 Type Ia SN model \citep{b19}, which is widely
used as the standard model.

\section{Discussion and Conclusions}

In this work, we provide for the first time a high-quality image,
spectra, and detailed description of the X-ray emission of
G346.6-0.2 using a set of public archival {\it Suzaku} data.  The
radio emission of this remnant extends to 8 arcmin \citep {b25},
while X-ray image shows a smaller extension (see Fig. 1), which is
consistent with that of the $ASCA$ results shown in \citet {b28}.

We fitted the spectra with ``thermal'' component in
non-equilibrium ionization and an additional ``non-thermal''
power-law component with reasonably good chi-square values.

\emph{Thermal emission.} The thermal part of the spectra can be
represented by non-equilibrium ionization (VNEI) model. The
thermal X-ray emission of young SNRs, whose emission is still
dominated by the ejecta, is predominantly from the ejecta heated
by the reverse shock. Since the ejecta is metal abundant, its
X-ray spectra usually shows strong emission lines of heavy
elements such as O, Ne, Mg, Si, S, Ar, Ca and Fe. $K\alpha$
emission lines of Mg, Si, S, Ca and Fe for G346.6-0.2 are detected
clearly for the first time in this work.

The interstellar absorbing column density $N_{\rm H}$ is obtained
to be ($2.1\pm 0.2$)$\times10^{22}$ $\rm cm^{-2}$, which is
consistent with the galactic value of ($1.5\pm
0.2$)$\times10^{22}$ $\rm cm^{-2}$ in that direction \citep {b6}.
$ASCA$ results of G346.6-0.2 reported a value of $N_{\rm
H}>10^{22}$ $\rm cm^{-2}$ \citep {b28}, indicating that X-ray
spectra of the remnant is heavily absorbed by interstellar matter,
perhaps due to its location being close to the Galactic plane. In
other words, the electron density ($n_{\rm e}$) of medium of the
remnant is high enough. This may also be supported by the fact
that the south region of the remnant is in interaction with four
molecular clouds as has been noted by \citet {b10}.

Being close to the Galactic plane and being in a highly dense
region, one can expect that some portions of the remnant can be
highly asymmetrical. The asymmetry in the south region of the
remnant may well be due to the interaction with the maser sources
as has been noted in Fig. 1. One can argue that the surface
brightness of the remnant is far from the equal distribution and
spherical symmetry due to the inhomogeneity of the medium.

The thermal emission further provides us information about the
electron temperature and age of the plasma. Our best-fitting
temperature is obtained to be $\sim$1.2 keV, which is typical for
shell-like SNRs. This value is also consistent with the result of
$\sim$1.6 keV obtained from {\it ASCA} observations \citep {b28}.

The age of G346.6-0.2 has not been predicted so far. Especially,
young SNRs whose ages are a few hundred or about a thousand years
are not only bright but also dominated by the emission from
ejecta. Being an ejecta-dominant remnant and  bright in X-rays
G346.6-0.2 can be a young SNR. Plasma in most of young SNRs are in
the NEI condition. The ionization age (or ionization parameter) is
defined as $\tau$=$n_{\rm e}t$, which is often used as a key
diagnostic of the NEI state. $\tau$ is typically required to be
$\geq$$10^{12}$ $\rm cm^{-3}$s \citep {b14} for full ionization
equilibrium. In our observations we obtained $n_{\rm e}t$ to be
(2.92$\pm0.01$)$\times10^{11}$ $\rm cm^{-3}$s, which may also
indicate that this SNR is far from full ionization equilibrium.
So, the plasma has not yet had time to reach ionization
equilibrium, and it is still being ionized. We calculated $n_{\rm
e}$ to be $\sim$0.82 $\rm cm^{-3}$ from VEM=$n_{\rm e}n_{\rm H}V$
where $n_{\rm e}=1.2n_{\rm H}$ and V is the X-ray-emitting volume
of the remnant (estimated to be $\sim$$2\times10^{54}$ $\rm
cm^{3}$) by adopting an average distances of d $\sim8.3$ kpc
(average of a near value of 5.5 kpc and a far value of 11 kpc
given by \citet {b10}) and assuming the emitting volume to be
spherical shell. From the equation t=$\tau$/$n_{\rm e}$, the age
of G346.6-0.2 calculated to be $\sim$$1.1\times10^{4}$ yr. Thus,
this remnant is likely to be the oldest known ejecta-dominated
shell-like SNR (e.g. SN1006, RCW86, G337.2-0.7 and G309.2-0.6,
$\sim$1000 yr \citep {b26}, $\sim$1800 yr \citep {b30}, 2000-4500
yr and 700-4000 yr \citep {b20}, respectively.)

\emph{Non-thermal emission.} Recently, in some SNRs (e.g. first
SN1006 \citep {b11} and then RX J1713.7.3946 \citep {b29, b34},
Cas A \citep {b32, b33}, and Tycho's SNR \citep {b31}) non-thermal
emission has been detected. There are two possible mechanisms for
non-thermal emission: one is synchrotron emission and the another
is non-thermal bremsstrahlung. The non-thermal emission coming
from G346.6-0.2 is most likely a synchrotron emission (power-law
component with photon index of $\Gamma$ $\sim 0.6$). The real
source of synchrotron emission in SNRs is believed to be the
acceleration of the ultrarelativistic electrons in the magnetic
fields with the help of the shock waves. A powerful X-ray point
source inside the remnant or in the vicinity of the remnant can
also contribute to the non-thermal emission. The photon index we
obtained ($\Gamma$ $\sim 0.6$) is lower (harder) than those of
typical shell-like SNRs. This forces us to presume that a pulsar
wind nebula (PWN) contribution is also possible, although a PWN
has not been discovered so far in or in the vicinity of
G346.6-0.2.

The flux value of G346.6-0.2 is obtained to be $F_{\rm x}$
$\sim4.4\times10^{-11}$ erg $\rm s^{-1}\rm cm^{-2}$ for $0.3-10$
keV range, which is in a good agrement with that of found with
$ASCA$ \citep {b28}. Using this flux value for the source and also
taking its X-ray angular size of $\sim$5 arcmin, its surface
brightness in the $0.3-10$ keV energy range is found to be
$\Sigma$ $\sim1.8\times10^{-12}$ erg $\rm s^{-1}\rm cm^{-2}\rm
arcmin^{-2}$, which is higher than many shell-like Galactic SNRs.
Surface brightness mainly depends on the explosion energy and the
density of the medium, but a PWN may also contribute to the
surface brightness of the SNR. In this case, high medium density
and small size of the remnant may be the main reason of high X-ray
surface brightness of G346.6-0.2.

\emph{Relative abundances in the ejecta. } Theoretical models
predict that in core collapse SN explosion low-Z elements like O,
Ne, Mg are produced \citep {b23}, while in Type Ia SNe \citep
{b19} high-Z elements like Ar, Ca and Fe are mostly produced.
Furthermore, Fe production in Type Ia SNe is far larger than that
of core-collapse SNe \citep {b19,b8}. The detection of Fe emission
lines from the remnants is important to identify the type of the
remnant. $Suzaku$ has detected Fe $K\alpha$ lines in especially
Type Ia SNRs (e.g. Tycho, SN1006, RCW86).

When we compare our best-fitting abundances of Mg, S, Ca and Fe
relative to Si with the predicted theoretical values, Mg relative
to Si is higher, while S and Ca relative to Si are lower. The
reason for the lower abundances of S and Ca could be because the
elements are concentrated at the inner layers of the remnant and
hence are not heated enough by the reverse shock yet. This
information implies that this SNR is at its early evolution phase.
Our best-fitting abundance Fe relative to Si is consistent with
the expected value (see Fig. 4) within the confidence range.
However, considering the large error bars due to low statistics,
we could also comment a similar statement for Fe abundance value
and an unattained reverse shock, as it is reported in many Type Ia
SNRs like SN1006 \citep {b26}, Tycho \citep {b27} and G337.2-0.7
\citep {b20} the Fe abundance is being less than expected value
produced in a Type Ia SNe. Considering all these cases we can
predict that the remnant may be originated from a Type Ia SN
explosion.

\section*{Acknowledgments}

A.S. is supported by T\"{U}B\.{I}TAK Post$-$Doctoral Fellowship.
This work is supported by the Akdeniz University Scientific
Research Project Management and by T\"{U}B\.{I}TAK under project
codes 108T226 and 109T092. The authors also acknowledge the
support by Bo\u{g}azi\c{c}i University Research Foundation under
2010-Scientific Research Project Support (BAP) project no:5052.

\begin{table*}
\centering
 \begin{minipage}{140mm}
  \caption{The best-fitting parameters of the G346.6-0.2 spectra in
$0.3-10$ keV energy band for VNEI, VMEKAL and VPSHOCK models.}
 \begin{tabular}{@{}cccc@{}}
  \hline
      Parameters & VNEI+power-law& VMEKAL+power-law&VPSHOCK+power-law \\
 \hline
 $N_{\rm H}$($\times10^{22}$ $\rm cm^{-2})$ & 2.1 $\pm 0.2$& 2.0 $\pm 0.2$& 2.1 $\pm 0.2$\\
$kT_{\rm e}$(keV) & 1.22$\pm 0.04$& 0.97$\pm 0.05$& 1.3$\pm 0.1$ \\
 $n_{\rm e}t $($\times10^{11}$$\rm cm^{-3}s$) & 2.92$\pm 0.01$& - & (6.92$\pm 0.01$)\footnote{The VPSHOCK fitting gives an upper limit on the ionization timescale.}\\
 Mg& 0.51$\pm 0.12$& 0.65$\pm 0.15$& 0.76$\pm 0.16$\\
 Si & 0.47$\pm 0.07$& 0.44$\pm 0.07$& 0.56$\pm 0.09$\\
 S & 0.7$\pm 0.1$& 0.7$\pm 0.1$& 0.8$\pm 0.1$\\
 Ca & 2.3$\pm 0.6$& 3.4$\pm 0.8$& 2.3$\pm 0.7$\\
Fe& 0.6$\pm 0.3$& 0.4$\pm 0.2$& 0.6$\pm 0.3$\\
VEM\footnote{Volume emission measure (VEM=$\int n_{\rm e}n_{\rm
H}$dV in the unit of $10^{58}$ $\rm cm^{-3}$, where $n_{\rm e}$
and $n_{\rm H}$ are number densities of electrons and protons,
respectively and V is
the X-ray-emitting volume.}&11.6$\pm 1.1$ & 15.1$\pm 1.9$&10.4$\pm 1.5$\\
Flux\footnote{Flux in the $0.3-10$ keV band in the unit of $10^{-11}$ erg $\rm s^{-1}$$\rm cm^{-2}$.} &3.1$\pm 0.5$ &2.9$\pm 0.6$  & 3.2$\pm 0.4$ \\
 Photon Index &0.6 $\pm 0.3$ &0.5 $\pm 0.3$&0.6 $\pm 0.3$\\
 norm($\times10^{-4}$photons $\rm cm^{-2}s^{-1}$) &$4.46\pm 0.04$ & 4.55$\pm 0.04$&4.55$\pm 0.05$\\
 Flux\footnote{Total flux in the $0.3-10$ keV band in the unit of $10^{-11}$ erg $\rm s^{-1}$$\rm cm^{-2}$.} &4.4$\pm 0.2$&4.5$\pm 0.2$ &4.4$\pm 0.2$ \\
 $\chi^{2}$/d.o.f.  &606/565=1.07 &607/566=1.07&605/565=1.07\\
\hline
\end{tabular}
\end{minipage}
\end{table*}
\begin{figure}
  \vspace*{17pt}
  \includegraphics[width=8cm]{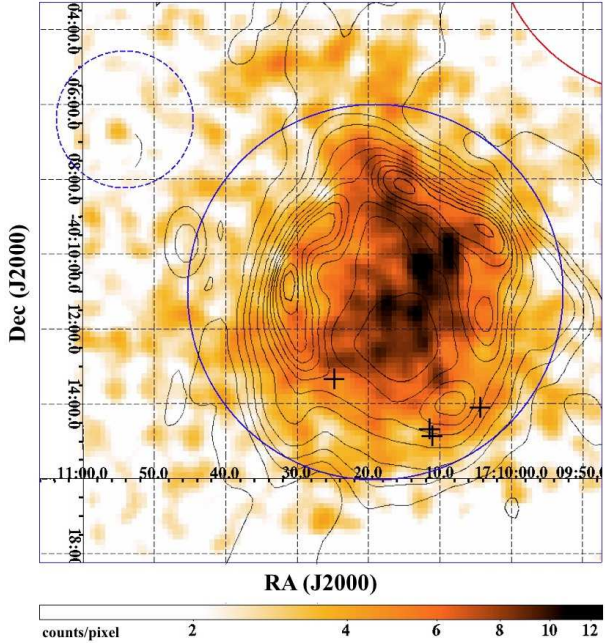}
  \caption{XIS0 image of G346.6-0.2 in the $0.3-10$ keV full energy
band. The coordinates are referred to epoch J2000. The calibration
source located at the upper right corner, as seen by red-circle,
is masked out. The colour-coding is for the X-ray photons in
counts/pixel. The overlaid iso-intensity contours are from radio
observation at 843 MHz. The contour levels are -7, 6, 25, 50, 95,
135, 165, 190, 210, 230, 255 and 270 mJy/beam, selected the same
as the radio works for visual comparison. The regions used to
extract spectrum and to determine the background parameters are
indicated by blue solid and dotted circles, respectively. The
positions of the four masers defined by \citet {b10} are marked as
crosses.} \label{sample-figure}
\end{figure}

\begin{figure}
\vspace*{17pt}
\includegraphics[width=7.7cm]{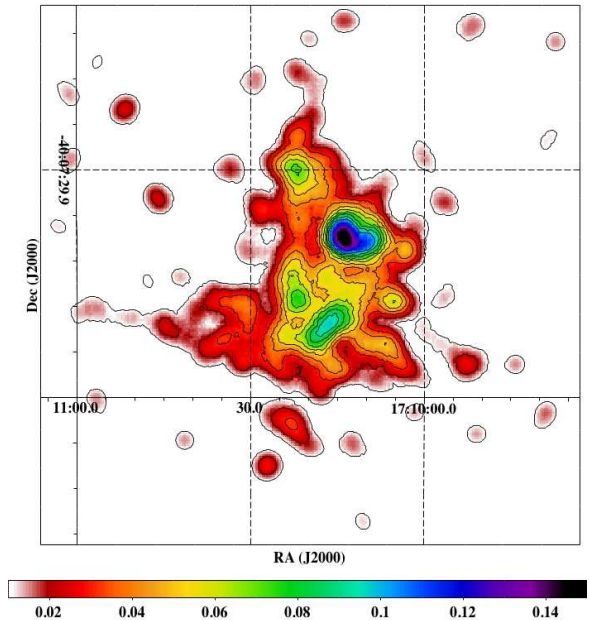}
\caption{G346.6-0.2 XIS0 image in $1.5-2.5$ keV (Si-S) energy
band. The image is smoothed for 10$\sigma$ Gaussian in order
highlight the Si-S distribution. Overlaid contours are spaced
linearly in intensity (0.01, 0.02, 0.03, 0.05, 0.09 and 0.16
counts/pixel).}
\end{figure}

\onecolumn
\begin{figure}
\centering
  \vspace*{13pt}
 \includegraphics[width=15cm]{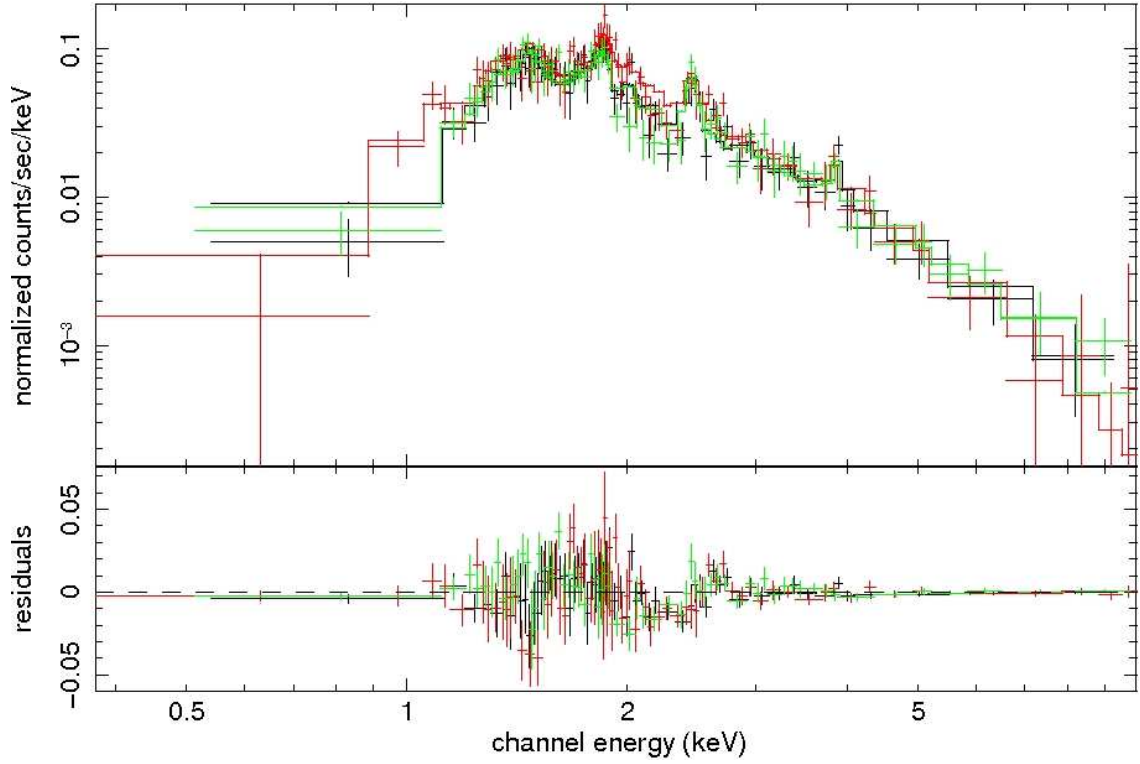}
  \caption{Background-subtracted XIS (XIS0:black, XIS1:red, XIS3:green) spectra of
 G346.6-0.2 in full energy band ($0.3-10$ keV) fitted with an
absorbed VNEI and power-law model. The lower panel shows the
residuals from the best-fitting model.}
\end{figure}
\twocolumn

\begin{figure}
  \vspace*{17pt}
  \includegraphics[width=8.5cm]{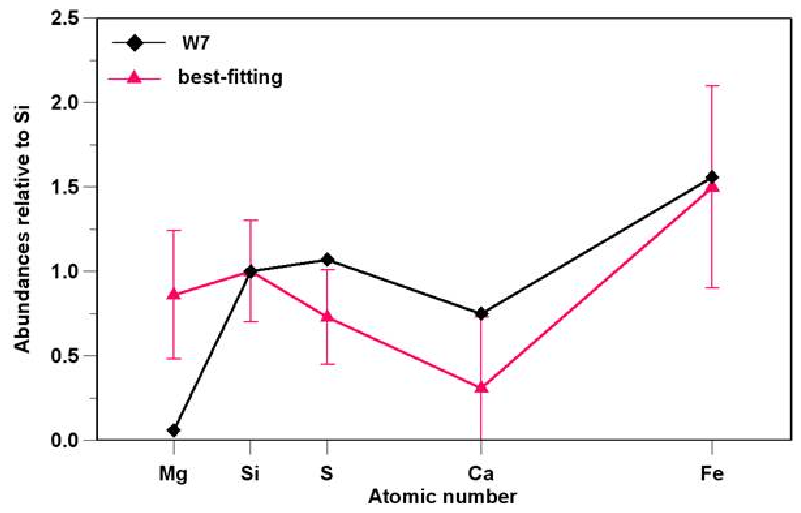}
  \caption{Metal abundances of Mg, S, Ca and Fe relative to solar values and
   normalized to Si. W7 model for a Type Ia is shown by diamonds while our data are shown by
   triangles.}
\end{figure}


\begin{thebibliography}{99}
\bibitem[\protect\citeauthoryear{Anders \& Grevesse}{1989}]{b1} Anders E., Grevesse N., 1989, Geochimica Cosmochimica Acta, 53, 197
\bibitem[\protect\citeauthoryear{Arnaud}{1996}]{b2} Arnaud K. A., 1996, in Jacoby G., Barnes J., eds, ASP Conf. Ser. Vol.101,
 Astronomical Data Analysis Software and Systems V. Astron.Soc. Pac., San Francisco, p. 17
\bibitem[\protect\citeauthoryear{Bamba, Koyama \& Tomida}{2000}]{b3} Bamba A., Koyama K., Tomida H., 2000, PASJ, 52, 1157
\bibitem[\protect\citeauthoryear{Borkowski, Lyerly \& Reynolds}{2001}]{b4} Borkowski K. J., Lyerly W. J., Reynolds S. P., 2001, ApJ, 548, 820
\bibitem[\protect\citeauthoryear{Clark, Caswell \& Green }{1975}]{b5} Clark D. H., Caswell J. L., Green A. J., 1975, Australian Journal of Physics, Astrophysical Supplement, 37, 1
\bibitem[\protect\citeauthoryear{Dickey \& Lockman}{1990}]{b6} Dickey J.M., Lockman F. J., 1990, ARA\&A, 28, 215
\bibitem[\protect\citeauthoryear{Hughes et al.}{2000}]{b33} Hughes J. P., Rakowski C. E., Burrows D. N., Slane P. O., 2000, ApJ, 528, L109
\bibitem[\protect\citeauthoryear{Hwang et al.}{2002}]{b31} Hwang U., Decourchelle A., Holt S. S., Petre R., 2002, ApJ, 581, 1101
\bibitem[\protect\citeauthoryear{Ishisaki et al.}{2007}]{b7} Ishisaki Y. et al., 2007, PASJ, 595, 113
\bibitem[\protect\citeauthoryear{Iwamoto et al.}{1999}]{b8} Iwamoto K., Brachwitz F., Nomoto K., Kishimoto N., Umeda H., Hix W. R., Thielemann F.-K., 1999, ApJS, 125, 439
\bibitem[\protect\citeauthoryear{Kokubun et al.}{2007}]{b9} Kokubun M. et al., 2007, PASJ, 59, 53
\bibitem[\protect\citeauthoryear{Koralesky et al.}{1998}]{b10} Koralesky B., Frail D. A., Goss W. M., Claussen M. J., Green A. J., 1998, AJ, 116, 1323
\bibitem[\protect\citeauthoryear{Koyama et al.}{1995}]{b11} Koyama K., Petre R., Gotthelf E. V., Hwang U., Matsuura M., Ozaki M., Holt S. S., 1995, Nat, 378, 255
\bibitem[\protect\citeauthoryear{Koyama et al.}{1997}]{b29}Koyama K., Kinugasa K., Matsuzaki K., Nishiuchi M., Sugizaki M., Torii K., Yamauchi S., Aschenbach B., 1997, PASJ, 49, L7
\bibitem[\protect\citeauthoryear{Koyama et al.}{2007}]{b12} Koyama K. et al., 2007, PASJ, 59, 23
\bibitem[\protect\citeauthoryear{Liedahl, Osterheld \& Goldstein }{1995}]{b13} Liedahl D. A., Osterheld A. L., Goldstein W. H., 1995, ApJ, 438, L115
\bibitem[\protect\citeauthoryear{Masai} {1984}]{b14} Masai K., 1984, Ap\&SS, 98, 367
\bibitem[\protect\citeauthoryear{Mewe, Gronenschild \& van den Oord}{1985}]{b15} Mewe R., Gronenschild E. H. B. M., van den Oord G. H. J., 1985, A\&AS, 62, 197
\bibitem[\protect\citeauthoryear{Mewe, Lemen \& van den Oord}{1986}]{b16} Mewe R., Lemen J. R., van den Oord G. H. J., 1986, A\&AS, 65, 511
\bibitem[\protect\citeauthoryear{Mitsuda et al.}{2007}] {b17} Mitsuda K. et al., 2007, PASJ, 59, 1
\bibitem[\protect\citeauthoryear{Morrison \& McCammon}{1983}]{b18} Morrison R., McCammon D., 1983, ApJ, 270, 119
\bibitem[\protect\citeauthoryear{Nomoto et al.}{1997}] {b19} Nomoto K., Iwamoto K., Nakasato N., Thielemann F.-K., Brachwitz F., Tsujimoto T., Kubo Y., Kishimoto N., 1997, Nuclear Physics A, 621, 467
\bibitem[\protect\citeauthoryear{Rakowski, Hughes \& Slane}{2001}] {b20}Rakowski C. E., Hughes J. P., Slane P., 2001, ApJ, 548, 258
\bibitem[\protect\citeauthoryear{Serlemitsos et al.}{2007}]{b21} Serlemitsos P. J. et al., 2007, PASJ, 59, 9
\bibitem[\protect\citeauthoryear{Slane et al.}{1999}]{b34}Slane P., Gaensler B. M., Dame T. M., Hughes J.P., Plucinsky P. P., Green A., 1999, ApJ, 525, 357
\bibitem[\protect\citeauthoryear{Takahashi et al.}{2007}]{b22} Takahashi T. et al., 2007, PASJ, 59, 35
\bibitem[\protect\citeauthoryear{Tamagawa et al.}{2009}]{b27} Tamagawa T. et al., 2009, PASJ, 61, 167
\bibitem[\protect\citeauthoryear{Thielemann, Nomoto \& Hashimoto}{1996}]{b23} Thielemann F-K., Nomoto K., Hashimoto M., 1996, ApJ, 460, 408
\bibitem[\protect\citeauthoryear{Vink et al.}{2000}]{b32} Vink J., Kaastra J. S., Bleeker J. A. M., Bloemen, H., 2000, Advances in Space Research, 25, 689
\bibitem[\protect\citeauthoryear{Warren et al.}{2005}]{b24} Warren J.S. et al., 2005, ApJ, 634, 376
\bibitem[\protect\citeauthoryear{Whiteoak \& Green} {1996}]{b25} Whiteoak J.B.Z., Green A.J., 1996, A\&AS, 118, 329
\bibitem[\protect\citeauthoryear{Yamaguchi et al.}{2008a}]{b26}Yamaguchi H. et al., 2008a, PASJ, 60, 141
\bibitem[\protect\citeauthoryear{Yamaguchi et al.}{2008b}]{b30}Yamaguchi H., Koyama K., Nakajima H.,
Bamba A., Yamazaki R., Vink J., Kawachi A., 2008b, PASJ, 60, 123.
\bibitem[\protect\citeauthoryear{Yamauchi et al.}{2008}]{b28} Yamauchi S., Ueno M., Koyama K., Bamba A., 2008, PASJ, 60, 1143

\end{thebibliography}
\end{document}